\documentclass[prl,twocolumn,superscriptaddress,amsmath]{revtex4-2}

\usepackage{graphicx}
\usepackage{braket}
\usepackage{xcolor}
\usepackage[normalem]{ulem}
\usepackage[bookmarks=false,colorlinks=true,urlcolor=blue,citecolor=blue,linkcolor=blue]{hyperref}

\begin{document}

\title{Dynamical entanglement transition in the probabilistic control of chaos}
\author{Thomas Iadecola}
\affiliation{Department of Physics and Astronomy, Iowa State University, Ames, IA 50011, USA}
\affiliation{Ames National Laboratory, Ames, IA 50011, USA}
\author{Sriram Ganeshan} 
\affiliation{Department of Physics, City College, City University of New York, New York, NY 10031, USA}
\affiliation{CUNY Graduate Center, New York, NY 10031}
\author{J. H. Pixley}
\affiliation{Department of Physics and Astronomy, Center for Materials Theory, Rutgers University, Piscataway, NJ 08854 USA}
\author{Justin H. Wilson}
\affiliation{Department of Physics and Astronomy, Louisiana State University, Baton Rouge, LA 70803, USA}
\affiliation{Center for Computation and Technology, Louisiana State University, Baton Rouge, LA 70803, USA}
\date{\today}

\begin{abstract}
    We uncover a dynamical entanglement transition in a monitored quantum system that is heralded by a local order parameter.
    Classically, chaotic systems can be stochastically controlled onto unstable periodic orbits and exhibit controlled and uncontrolled phases as a function of the rate at which the control is applied.
    We show that such control transitions persist in open quantum systems where control is implemented with local measurements and unitary feedback.
    Starting from a simple classical model with a known control transition, we define a quantum model that exhibits a diffusive transition between a chaotic volume-law entangled phase and a disentangled controlled phase. 
    Unlike other entanglement transitions in monitored quantum circuits, this transition can also be probed by correlation functions without resolving individual quantum trajectories.
\end{abstract}

\maketitle

The dynamics of quantum many-body systems host a range of phenomena usually out of reach to the classical world. 
Of particular relevance is the ability to locally measure and control systems to enable quantum technological applications such as efficient state preparation~\cite{foss-feig_holographic_2020,tantivasadakarn_long-range_2022,lu_measurement_2022,Friedman22}, quantum error correction~\cite{gottesman_introduction_2009,nielsen_quantum_2011}, and non-destructive measurements~\cite{hurst_measurement-induced_2019,hurst_feedback_2020}. 
Allowing such nonunitary ``hybrid" dynamics in the evolution of many-body states enriches the problem beyond simple, unitary evolution alone. This has recently led to the discovery of phases and phase transitions in the entanglement structure due to the competition between entangling unitary dynamics and projective local measurements~\cite{li_quantum_2018,li_measurement-driven_2019,skinner_measurement-induced_2019,potter_entanglement_2021}. 

The measurement-induced phase transition (MIPT) in its original formulation entails a fundamental change in the scaling of entanglement from volume- to area-law that is connected to percolation~\cite{skinner_measurement-induced_2019,vasseur_entanglement_2019,bao_theory_2020}, but it has grown past that paradigm and has been studied in numerous contexts~\cite{gullans_dynamical_2020,jian_measurement-induced_2020,li_conformal_2021,zabalo_critical_2020,zabalo_operator_2022,lavasani2021measurement,bao_symmetry_2021,li2021robust,agrawal_entanglement_2021,barratt_field_2021,ippoliti_entanglement_2021,lavasani_topological_2021,van_regemortel_entanglement_2021,lang_entanglement_2020,sang_measurement-protected_2021,choi_quantum_2019,alberton_entanglement_2021,szyniszewski_entanglement_2019,bao_theory_2020,lunt_measurement-induced_2020,goto_measurement-induced_2020,tang_measurement-induced_2020,cao_entanglement_2019,nahum_measurement_2021,turkeshi_measurement-induced_2020,zhang_nonuniversal_2020,szyniszewski_universality_2020,fuji_measurement-induced_2020,rossini_measurement-induced_2020,vijay2020measurement,turkeshi_measurement-induced_2021,sierant2022dissipative,sharma2022measurement,chen2021non,han_measurement-induced_2022,iaconis_measurement-induced_2020,buchhold_effective_2021,ladewig_monitored_2022,muller_measurement-induced_2022,altland_dynamics_2022, willsher_measurement-induced_2022, barratt_transitions_2022,Cote22}.
While several incarnations of the transition have been demonstrated, this entanglement transition can only be witnessed by quantities that are non-linear in the density matrix; correlation functions averaged over measurement outcomes are unaffected by the local measurements in the long-time limit. 
This makes observing the MIPT in experiment a significant challenge requiring either postselection or decoding.

If, on the other hand, we augment each local measurement with control \cite{mcginley_absolutely_2021-1,herasymenko_measurement-driven_2021} (i.e., unitary feedback conditioned on the measurement outcome), it should be possible to stabilize a dynamical phase transition that is observable in quantities that are linear in the density matrix. In this work, we identify such a control transition in an open quantum many-body system. Unlike previously studied MIPTs, incorporating local feedback into the dynamics leads to a unique control transition visible in both entanglement measures and correlation functions, making it observable using current experimental setups.

The central idea stems from classical dynamical systems, where methods to control chaotic dynamics have been developed. 
The control protocols can be either deterministic, requiring constant monitoring and perturbation \cite{ott1990controlling}, or probabilistic \cite{antoniou_probabilistic_1996, antoniou_absolute_1998}.  
The latter entails the coupled stochastic action of a chaotic map (with probability $1-p$) together with a control map (with probability $p$). 
These two coupled maps have the same periodic orbit, unstable for the chaotic map and stable for the control map. 
Under the stochastic action of the coupled map, the periodic orbit becomes the global attractor at some critical control rate $p_{\rm ctrl}$. 
\emph{Prima facie}, the control transition in these maps seems to mirror certain aspects of MIPTs, albeit at a purely classical level, with the control map as a classical proxy for quantum measurements. The question then naturally arises of whether we can construct a quantum version of probabilistic control transitions and contrast these with quantum MIPTs.

\begin{figure}
    \centering
    \includegraphics[width=.95\columnwidth]{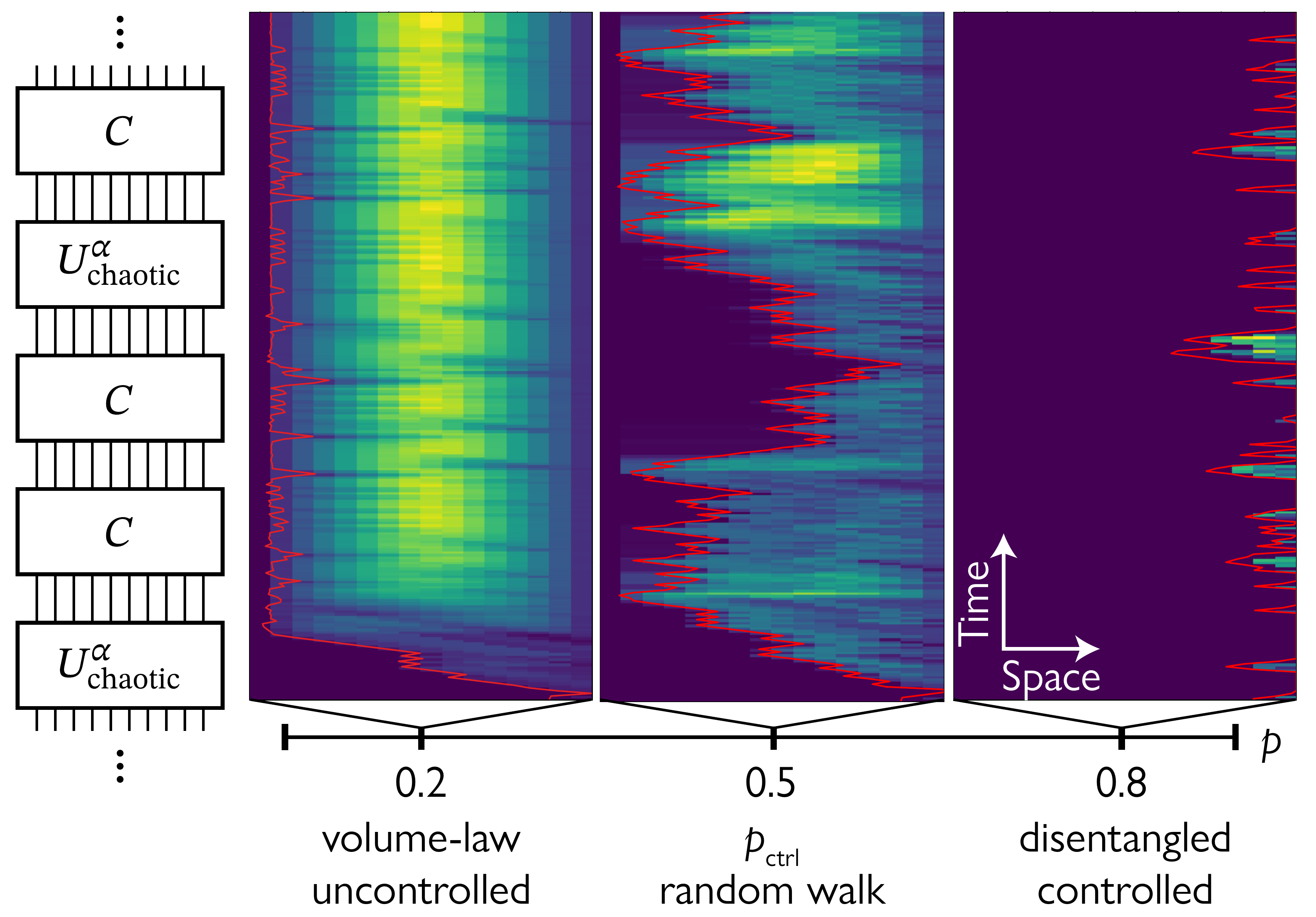}
    \caption{\textbf{Control transition.} 
    The quantum transition can be seen in the dynamics of single trajectories, shown above for $L=16$. (left) A particular realization of the stochastic quantum circuit composed of unitary dynamics $U_{\mathrm{chaotic}}^{\alpha}$ and a nonunitary control map $C$. (right) The phase diagram as a function of the control probability $p$ (black horizontal line). For $p< p_\mathrm{ctrl}$, the domain wall between controlled and uncontrolled regions (red) is swept to the left, and entanglement (color scale with brighter colors indicating higher entanglement) grows behind it in a volume-law fashion. For $p>p_{\rm ctrl}$, the domain wall is pinned to the right edge, and the region to its left is fully controlled and disentangled. At $p = p_\mathrm{ctrl}$, the domain wall undergoes an unbiased random walk and the entanglement typically grows to an $O(1)$ value.
    }
    \label{fig:Entanglement-DW}
\end{figure}

In this paper, starting from a classically chaotic map with a known control transition, we construct a suitable quantum model involving measurements and feedback in which the control transition is enriched by quantum entanglement. 
The phase transition in the quantum model can be probed by an appropriate local order parameter, and is diffusive with dynamic exponent $z \approx 2$ and correlation length exponent $\nu\approx 1$, similar to the classical case. 
However, the transition is also observed in entanglement and purification measures traditionally used to diagnose MIPTs, which have no analogue in the classical limit. 
In contrast to the feedback-free MIPTs, the entanglement entropy grows diffusively at the transition before saturating to an area-law value. 
Many properties of the transition can be understood by tracking the dynamics of an emergent semiclassical domain wall, which undergoes an unbiased random walk at the transition, see Fig.~\ref{fig:Entanglement-DW}.

\emph{Model.}
One of the simplest classical models with a control transition is the Bernoulli map~\cite{renyi1957representations}. 
This map operates on the phase space $[0,1)$ and is given by 
\begin{equation}
    x \mapsto 2x \!\!\mod 1. \label{eq:classical_bernoulli}
\end{equation}
It has a Kolmogorov-Sinai entropy of $\log 2$~\cite{renyi1957representations}.
In this chaotic map, any rational number $x_0 = a/b$ such that 2 does not divide $b$ undergoes a finite-length periodic orbit.
For instance, if $x_0 = 1/3$, then $x_1 = 2/3$ and $x_2 = 1/3 = x_0$; this is a periodic orbit of length 2.
However, since the rational numbers are a set of measure zero in the interval $[0,1)$, almost every initial state in the interval undergoes chaotic dynamics.
To control this dynamics onto a periodic orbit of our choosing (with points $\{x_j\}$), we define connected regions $\Delta_j$ such that $x_j \in \Delta_j$ and $\cup_j \Delta_j = [0, 1)$, yielding the control map \cite{antoniou_probabilistic_1996}
\begin{equation}
    x \mapsto  (1-a) x_j+ a x \quad\mathrm{if}\quad x \in \Delta_j. \label{eq:control}
\end{equation}
Note that the points 
$x_j$ are attractive fixed points of the control map for $|a|<1$.
We then consider a discrete-time stochastic dynamics in which, at each time step, the chaotic map \eqref{eq:classical_bernoulli} is applied with probability $1-p$ and the control map is applied with probability $p$. 
For a critical control rate $p_{\rm ctrl}$, there is a phase transition between an uncontrolled phase in which the system never reaches the periodic orbit and a controlled phase in which the system always reaches the orbit.
For each value of $|a| < 1$ in Eq.~\eqref{eq:control}, there is a different $p_\mathrm{ctrl}$~\cite{antoniou_absolute_1998}.

To rephrase this model in a manner compatible with quantum mechanics, we map it to a system of qubits as follows. Write the point $x\in[0,1)$ in base 2 as $x = 0.b_1 b_2 b_3 \cdots$ where $b_i \in \{0,1\}$. 
Then, $2x \mod 1 = 0.b_2 b_3 b_4 \cdots$, or in other words, Eq.~\eqref{eq:classical_bernoulli} implements a leftward shift of the bitstring.
Then it is natural to define a Hilbert space spanned by computational basis (CB) states $\ket{b_1 b_2 b_3\cdots} \equiv \ket{x}$.
To make the quantum problem more tractable, we truncate the Hilbert space to consist of bitstrings of length $L$.
Upon doing this, we immediately encounter a problem: this map appears to be nonunitary (it erases $b_1$), and there is an ambiguity as to the value of $b_L$.
The first solution to this problem is simple: let $b_L \equiv b_1$ such that the Bernoulli map becomes the translation operator
\begin{equation}
    T \ket{b_1 b_2 \cdots b_L} = \ket{b_2 b_3 \cdots b_L b_1}. \label{eq:translation}
\end{equation}
However, in this formulation, every initial state belongs to a periodic orbit of length at most $L$.
In order to have a notion of chaos in the thermodynamic limit $L\to\infty$, we need the typical orbit length to be exponential in $L$~\cite{Takesue89,Gopalakrishnan18,Iadecola20}. 
To accomplish this, we compose the map in Eq.~\eqref{eq:translation} with a scrambling operation on the last few qubits.
We consider two options for the scrambling operation: a ``classical" and a ``quantum" one.
The former acts as a permutation on the 8-dimensional space of bitstrings $b_{L-2}b_{L-1}b_L$, while the latter is a Haar-random unitary acting on the last two qubits.
The chaotic unitary is then defined as 
\begin{equation}
U_\mathrm{chaotic}^\alpha = S_\alpha T
\end{equation}
where $S_\alpha$ is the scrambler and $\alpha = \mathrm{qm}$ or $\mathrm{cl}$ for quantum or classical, respectively.
The unitary map $U^{\rm cl}_{\rm chaotic}$ is classical in the sense that it maps CB states to CB states---it is a reversible cellular automaton.
We choose $S_{\rm cl}$ such that $U^{\rm cl}_{\rm chaotic}$ is a chaotic cellular automaton with typical orbit length $e^{O(L)}$ [see supplemental material (SM)].
In contrast, $U^{\rm qm}_{\rm chaotic}$ has no obvious notion of orbits 
and generates chaotic quantum dynamics in the sense that an initial CB state develops volume-law entanglement in an $O(L^2)$ time owing to the locality of the scrambler $S_{\rm qm}$ (see SM).

We implement the (inherently nonunitary) control map via measurement and feedback. 
To build it, we choose the period-2 orbit $\{x_0 = 1/3, x_1 = 2/3\}$ and set $a = 2^{-1}$ in Eq.~\eqref{eq:control}. 
For this value of $a$, the classical control transition is known to occur at $p_{\rm ctrl}=0.5$~\cite{antoniou_absolute_1998}.
It is then useful to break up the control map into the multiplication $\ket{x}\mapsto \ket{2^{-1} x}$ followed by the the addition $\ket{2^{-1} x}\mapsto\ket{2^{-1} x+2^{-1} x_j}$.
The full control operation is
\begin{equation}
    C = A_\mathrm{ctrl} T^{-1} R_L.
\end{equation}
$T^{-1}R_L$ implements the multiplication as follows.
$R_L$ projectively measures qubit $L$ in the CB and flips it if the outcome is $\ket{b_L}=\ket{1}$:
\begin{equation}
    R_L \ket{\psi} =  \begin{cases}
    \frac{ P^0_{L} \ket{\psi}}{\left\| P^0_{L} \ket{\psi} \right\|} & \text{with probability }\left\| P^0_{L} \ket{\psi} \right\|^2, \\
    \frac{ X_L P^1_{L} \ket{\psi}}{\left\| P^1_{L} \ket{\psi} \right\|} & \text{with probability }\left\| P^1_{L} \ket{\psi} \right\|^2, \\
    \end{cases}
\end{equation}
where $P^0_L$ and $P^1_L$ project the $L$th qubit onto $\ket{0}$ and $\ket{1}$, respectively, and $X_L$ is the Pauli-$X$ operator at site $L$ that sets $b_L=0$ if we measure $b_L=1$.
Subsequently, $T^{-1}\ket{ b_1 b_2 \cdots b_{L-1} 0} = \ket{0 b_1 b_2 \cdots b_{L-1}}$ completes the multiplication operation. Next, we apply a controlled adder circuit $A_\mathrm{ctrl}$ which acts on $\ket{x}=\ket{b_1 b_2 \cdots b_L}$ as
\begin{align}
A_{\rm ctrl}\ket{x}
=
\begin{cases}
\ket{x+0.00101\cdots011} & \text{if $b_2=0$}\\
\ket{x+0.01010\cdots101} & \text{if $b_2=1$}
\end{cases}
.
\end{align}
$A_{\rm ctrl}$ can be built from local unitary operations as described in the SM. 
Conditioning the addition on the value of $b_2$ determines whether the control map pushes the initial state $\ket{x}$ toward $\ket{x_0}$ or $\ket{x_1}$.

Putting everything together, the quantum model applies $U^\alpha_\mathrm{chaotic}$ with probability $1-p$ and $C$ with probability $p$ at a given time step. 
When the chaotic dynamics is generated by $U^{\rm cl}_{\rm chaotic}$, this stochastic dynamics occurs in the space of CB states, and is therefore equivalent to a probabilistic cellular automaton whose properties vary with $p$. 
The dynamics in this limit is classical, even though the model is phrased quantum mechanically. 
When $U^{\rm cl}_{\rm chaotic}$ is replaced by $U^{\rm qm}_{\rm chaotic}$, the chaotic dynamics becomes entangling, while $C$ disentangles the system by pushing it towards the periodic orbit of the underlying classical model.
These dynamics can also be formulated as a quantum channel; with additional dephasing, the superoperator that evolves the average density matrix reduces to the Frobenius-Perron evolution operator for classical phase space distributions (see SM).

\begin{figure}
    \centering
    \includegraphics[width=\columnwidth]{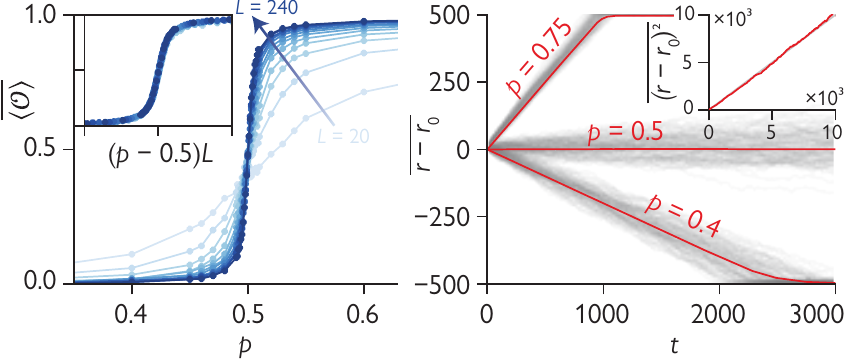}
    \caption{\textbf{Classical transition.} (left) Realization-averaged order parameter $ \overline{\langle \mathcal O \rangle} $ for various system sizes. 
    The crossing near $p = 0.500(1)$ is where the transition occurs. 
    (Inset) Collapse indicates that $\nu = 1.00(2)$. 
    (right) The position of the FDW initialized at $L/2$ for $L=1000$ in the controlled phase ($p=0.75$), the uncontrolled phase ($p=0.4$) and at the transition ($p=0.5$). 
    Gray curves are all 1000 realizations and red curves are averages. 
    (Inset) At $p=0.5$ we see random walk behavior in four orders of magnitude for $r^2 = t$ and a fit confirms $z = 2.04(8)$.
    } 
    \label{fig:classical_data}
\end{figure}

\emph{Classical transition.}
Our first order of business is to show that the classical control transition survives the above mapping to qubits.
To characterize the transition, we first note that the orbit $\{x_0 = 1/3, x_1 = 2/3\}$ is a two-dimensional subspace spanned by the CB states $\ket{1/3}=\ket{0101\cdots01}$ and $\ket{2/3}=\ket{1010\cdots10}$.
Thus, the control map \eqref{eq:control} steers the system's dynamics onto N\'eel-ordered antiferromagnetic states.
We probe this order using the order parameter
\begin{equation}
    \mathcal O = - \frac1{L} \sum^{L}_{i = 1} Z_i Z_{i+1}, \quad Z_{L+1} \equiv Z_1,
\end{equation}
where $Z_i$ is the Pauli $Z$ operator for bit $i$ [$Z_i\ket{b_i} = (-1)^{b_i+1}\ket{b_i}$].
The two N\'eel states maximize $\braket{\mathcal O} = 1$, so the controlled phase can be viewed as an ordered phase characterized by $\braket{\mathcal O}\to 1$ in the thermodynamic limit.
To probe the transition into the ordered phase in the classical case, we simulate the dynamics of CB states under the stochastic action of $U^{\rm cl}_{\rm chaotic}$ and $C$ out to $2L^2$ time steps for a range of of $p$ and $L$. 
For each $p$ and $L$, we calculate $\braket{\mathcal O}$ at the final time, and average the result over 1000 randomly chosen initial states and circuit instances.
We refer to this realization-averaged quantity as $\overline{\braket{\mathcal O}}$.
Our results, shown in Fig.~\ref{fig:classical_data}(left), show that N\'eel order develops for $p\gtrsim 0.5$. 
Scaling collapse with an ansatz $\overline{\braket{\mathcal O}} = f(L^{1/\nu}(p-p_{\rm ctrl}))$ is consistent with a transition point $p_{\rm ctrl}= 0.500(1)$, coinciding with the known result for Eqs.~\eqref{eq:classical_bernoulli} and \eqref{eq:control}~\cite{antoniou_absolute_1998}, and with a correlation length critical exponent $\nu= 1.00(2)$.
Further, the fluctuations of $\braket{\mathcal O}$ over realizations peak at $p=0.5$ (not shown), thereby serving as another indicator of the transition.

To further characterize the dynamics at the control phase transition, it is useful to consider the behavior of the ``first'' (i.e., leftmost) N\'eel domain wall in the chain, namely the local motif $11$ or $00$. 
The position of the first domain wall (FDW) bounds the distance from a point $x\in [0,1)$ to the periodic orbit: if the FDW is located on the $r$th bond in the chain, then $\text{min}_j|x-x_j|\lesssim O(2^{-r})$ for $j=0,1$.
The FDW thus constitutes the boundary between controlled and uncontrolled regions of the qubit chain, see Fig.~\ref{fig:Entanglement-DW}.
We simulate the dynamics of the FDW when initialized at $r_0=L/2$ and find averaged displacement and mean-squared displacement consistent with a random walk with bias $2p-1$ [see Fig.~\ref{fig:classical_data}(right)]:
\begin{equation}
    \overline{\braket{r-r_0}} = (2p - 1) t, \quad \overline{\braket{(r-r_0)^2}}\rvert_{p=0.5} = t.
\end{equation}
Fitting $\overline{\braket{(r-r_0)^2}}\sim t^{2/z}$ at $p=0.5$ confirms that $z=2.04(8)$, consistent with an unbiased random walk.

\begin{figure*}
    \centering
    \includegraphics[width=\textwidth]{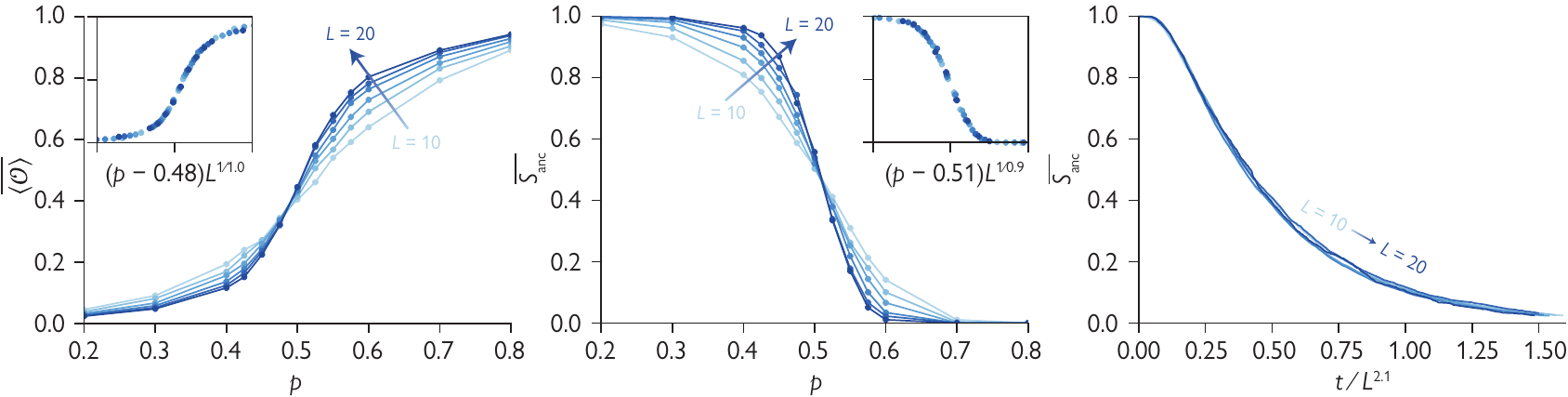}
    \caption{{\bf Quantum transition.} (left) Order parameter $\braket{\mathcal O}$ at time $2L^2$ averaged over initial states and circuit realizations. Inset shows scaling collapse assuming $\nu=1.0(1)$ and $p_{\rm ctrl}\approx 0.48(1)$. (middle) Ancilla entanglement order parameter $S_{\rm anc}$ (in units of $\ln 2$) at time $L^2/2$ averaged over initial states and circuit realizations. Inset shows scaling collapse assuming $\nu=0.9(1)$ and $p_{\rm ctrl}\approx 0.51(1)$. (right) Dynamics of $\overline{S_{\rm anc}}$ collapsed as a function of rescaled time $t/L^z$ near the control transition ($p=0.5$) with $z=2.1(1)$. 
    Data are averaged over $2000$ realizations for $L=10,\dots,16$, and $1000$ realizations for $L=18,20$. (For $\overline{S_{\rm anc}}$ data, $500$ realizations are used for $L=20$.) All points have error bars indicating standard error of the mean; where not visible, they are smaller than the points.} 
    \label{fig:quantum-transition}
\end{figure*}

\emph{Quantum transition.}
Next, we examine the fate of the control transition when the classical scrambler $S_{\rm cl}$ is replaced by the Haar-random scrambler $S_{\rm qm}$. Since the hybrid control circuit \eqref{eq:control} distributes over superpositions of CB states, a natural hypothesis is that the control transition survives. Then, above some critical $p$, the control circuit drives the system to a disentangled state with $\braket{\mathcal O}\to 1$ as $L\to\infty$, while below this critical value the system enters a volume-law entangled steady-state with $\braket{\mathcal O}\to 0$ as $L\to\infty$. Thus, in addition to a control transition, we expect to see an entanglement transition along the lines of those encountered in feedback-free MIPTs, but governed by a distinct universality class.

Exact numerical results bear out this simple picture. 
Fig.~\ref{fig:quantum-transition}(left) shows the realization-averaged order parameter $\overline{\braket{\mathcal O}}$ as a function of $p$. $\braket{\mathcal O}$ is measured at time $t=2L^2$, sufficiently late to ensure that the system reaches a steady state for all values of $p$ considered (see SM).
Similar to the classical case, there is a clear crossing near $p=0.5$. The inset of Fig.~\ref{fig:quantum-transition}(left) shows a scaling collapse assuming $\nu=1.0(1)$ and $p_{\rm ctrl}=0.48(1)$, suggesting a control transition near the expected location. 
In the SM, we show that the fluctuations over realizations of $\braket{\mathcal O}$ peak at $p=0.5$, similarly to the classical transition, and also collapse with $\nu\approx 1$. 

We further investigate the nature of this transition using tools developed for MIPTs. One perspective on MIPTs is that they are purification transitions above which an initially mixed density matrix becomes pure in a finite time~\cite{gullans_dynamical_2020}. This purification transition can be probed by preparing the system in a maximally entangled state with one ancilla qubit and tracking the ancilla's entanglement entropy, $S_{\rm anc}$, as a function of time~\cite{gullans_scalable_2020} for varying $L$ and $p$. At the purification transition, we expect a crossing of the $\overline{S_{\rm anc}}$-vs.-$p$ curves for different $L$ at times of order $L^2$. Fig.~\ref{fig:quantum-transition}(middle) shows such a crossing near $p=0.5$; the inset shows data collapse assuming $\nu=0.9(1)$ and $p_{\rm ctrl}=0.51(1)$. These data are taken after evolving the system for a time $t=L^2/2$, but the results are insensitive to small variations of this hyperparameter. To characterize the quantum dynamics at the transition, we consider $\overline{S_{\rm anc}}(t)$ in Fig.~\ref{fig:quantum-transition}(right) at $p=0.5$. We find that the curves for various $L$ nearly collapse upon rescaling $t\to t/L^z$ with $z=2.1(1)$, consistent with the dynamical exponent of the classical transition.

\begin{figure}
    \centering
    \includegraphics[width=\columnwidth]{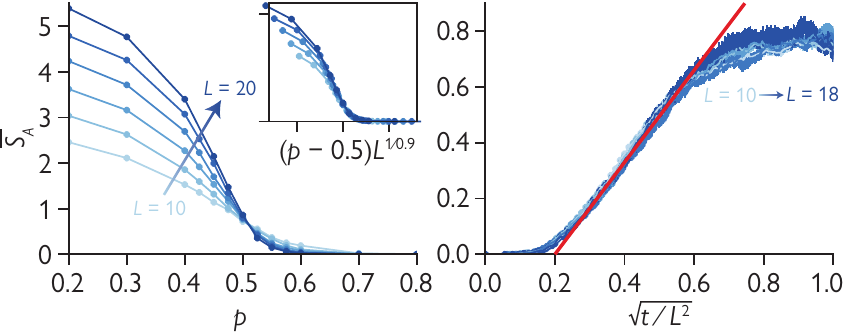}
    \caption{{\bf Entanglement structure and dynamics.} (left) Realization-averaged von-Neumann entanglement entropy $\overline{S_A}$ at time $2L^2$ for various $L$ and $p$. At large $p$ $\overline{S_A}$ decreases with $L$, while at small $p$ it increases linearly with $L$. There is a crossing near $p=0.5$, suggesting area-law entanglement at the transition.  Inset: Scaling collapse of $\overline{S_A}$ assuming $\nu=0.9(1)$ and $p_{\rm ctrl}=0.50(1)$. The data collapse near and above the control transition. (right) $\overline{S_A}$ as a function of rescaled time $\sqrt{t/L^2}$ near the quantum transition ($p=0.5$). The entanglement dynamics nearly collapse, and there is an intermediate-time regime where $\overline{S_A}\sim\sqrt{t}$ (red line).} 
    \label{fig:entanglement}
\end{figure}

Another perspective on MIPTs is that they constitute a volume-to-area-law transition in the entanglement entropy of a pure state. 
In Fig.~\ref{fig:entanglement}, we show that the system's entanglement entropy is also sensitive to the control transition. We calculate the von-Neumann entanglement entropy of the half-chain, $S_A$, taking region $A$ to be the leftmost $L/2$ sites of the chain. In Fig.~\ref{fig:entanglement}(left) we show $\overline{S_A}$ as a function of $p$ for different $L$, finding that it increases with $L$ for $p\lesssim 0.51$ and decreases with $L$ for $p \gtrsim0.51$. At the transition, we find that the wavefunction is area-law entangled on average, as indicated by a data collapse (inset) assuming $\nu=0.9(1)$ and $p_{\rm ctrl}=0.50(1)$. In Fig.~\ref{fig:entanglement}(right), we plot $\overline{S_A}(t)$ for $L=10,\dots,18$. The results collapse as a function of $\sqrt{t/L^2}$ (see also Fig.~\ref{fig:quantum-transition}) consistent with the classical expectation. 
In the early-time regime $t\ll L$, the realization-averaged entanglement grows diffusively, $\overline{S_A}(t,L)\sim \sqrt{t}/L$.

The entanglement properties at the transition can also be understood using the FDW. In the quantum model, the FDW becomes a wavepacket with average position $\braket{r(t)}=\sum_x\lvert\braket{x|\psi(t)}\rvert^2\, r_x$, where $r_x$ is the position of the FDW in the CB state $\ket{x}$. In the quantum setting, the uncontrolled region to the right of the FDW develops entanglement due to the action of the local scrambler $S_{\rm qm}$ (see Fig.~\ref{fig:Entanglement-DW}). The FDW thus constitutes a front between entangled and disentangled regions, and its dynamics governs that of the half-cut entanglement. At the transition the transport of the FDW is diffusive, so the entanglement dynamics must also be diffusive. Futhermore, since volume-law entanglement can only develop when the FDW ``sticks" to the left edge of the chain for at least an $O(L)$ time (see SM), which is exponentially unlikely in an unbiased random walk, the average entanglement can be at most area-law at the transition.

\emph{Discussion and outlook.}
In this work, we construct a quantum model that generalizes the stochastic dynamics associated with the probabilistic control of classical chaos. The model exhibits a dynamical entanglement transition reminiscent of MIPTs, but which is also witnessed by a local order parameter. Taken together, our numerical results for the quantum model indicate a diffusive transition at a critical control rate $p_{\rm ctrl}=0.5\pm 0.02$, consistent with both large-system numerics and previous analytical results for the classical version of the transition. The qualitative features of the classical transition are robust to quantum effects, but the entanglement dynamics across the transition has no classical analogue.

A natural question arising from this work is whether other MIPTs can be enriched into control transitions by the addition of feedback (e.g., contingent on what could be learned from the measurement record \cite{barratt_transitions_2022}). If this can be achieved, then the local order parameter for the control transition becomes an indicator for the entanglement transition. An immediate consequence is that measuring the order parameter for the controlled phase becomes sufficient to establish the phase transition experimentally, without requiring postselection onto individual quantum trajectories. We therefore expect that quantum control transitions of the type proposed here can be observed in a variety of noisy intermediate-scale quantum experiments.

\begin{acknowledgments}
\emph{Acknowledgements.}
J.H.W. and S.G. thank Penstock Coffee Shop in Highland Park, NJ where the seeds of this project were germinated.
We thank Piotr Sierant for helpful comments on an earlier version of the manuscript.
J.H.W. acknowledges illuminating discussions with Michael Buchhold.
This work was supported in part by the National Science Foundation under grants DMR-2143635 (T.I.) and OMA-1936351 (S.G.).
J.H.P.~was supported by the Alfred P. Sloan Foundation through a Sloan Research Fellowship.
This work was initiated, performed, and completed at the Aspen Center for Physics, which is supported by National Science Foundation grant PHY-1607611.
\end{acknowledgments}

\bibliography{refs, zotero_refs}

\pagebreak
\widetext
\clearpage

\setcounter{equation}{0}
\setcounter{figure}{0}
\setcounter{table}{0}
\setcounter{page}{1}
\renewcommand{\theequation}{S\arabic{equation}}
\setcounter{figure}{0}
\renewcommand{\thefigure}{S\arabic{figure}}
\renewcommand{\thepage}{S\arabic{page}}
\renewcommand{\thesection}{S\arabic{section}}
\renewcommand{\thetable}{S\arabic{table}}
\makeatletter

\textbf{Supplemental Material: Dynamical entanglement transition in the probabilistic control of chaos}

\section{Random walker description of $\beta$-adic map under control}

In this section of the supplementary, we provide a short review of the coupled stochastic dynamics pertaining to the probabilistic control of chaos~\cite{antoniou_probabilistic_1996, antoniou_probabilistic_1997, antoniou_absolute_1998}. We will also provide a random walk picture of the $\beta$-adic map in the presence of the control. The $\beta=2$ case considered in the main text is the special case and is also known as the Bernoulli map. The $\beta$-adic Renyi map can be written as,
\begin{align}
    R(x)=\beta x\quad \text{mod}\quad 1. \label{eq:supp_bern}
\end{align}
The above map has unstable periodic orbits of varying periods for every rational number initialization. On the other hand, the control map is designed such that a given unstable orbit is an attractive fixed point for $C(x)$,
\begin{align}
    C(x)=\sum_{j=1}^{N} (a x +(1-a) x_j ) 1_{\Delta_j}(x),\quad |a|<1 \label{eq:supp_ctrl}
\end{align}
 The $C(x)$ acts within a slice of the phase space defined by the step function $1_{\Delta_j}(x)$ containing the periodic orbit; $x_j$ are the $N$ points of the periodic orbits which are the attractive fixed points owing to the relations $C(x_j)=x_j$ and $|C'(x_j)|<1$.
 The collective stochastic dynamics of this map is the central premise of the classical probabilistic control of chaos.
 In order to quantify the coupled stochastic dynamics, we use the Frobenius-Perron (FP) operator.
 The FP operator acts on the ensembles of trajectories described by a probability distribution $\rho(x)$ as $U \rho_n(x)=\rho_n(M^{-1}(x))$.
 The FP operator for the total coupled dynamics can be written as,
 \begin{align}
     U_\mathrm{total}\rho(x)=(1-p)\frac{1}{\beta}\sum_{r=0}^{\beta-1}\rho\left(\frac{x+r}{\beta}\right)+p\sum_{j=1}^N \rho\left(\frac{ x-(1-a)x_j}{a}\right)1_{C(\Delta_j)}(x)\label{eq:FPtotal}
 \end{align}
 For the above map, the absolute controllability conditions have been derived in Ref.~\cite{antoniou_absolute_1998} and the critical rate of control dynamics has been shown to be,
 \begin{align}
   p_c=\frac{\log\beta}{\log \beta-\log a}  \label{eq:pctrl}
 \end{align}
 For the case considered in the main text $\beta=2, a=1/2$, we get $p_c=1/2$ consistent with our numerical results.

Eq.~\eqref{eq:pctrl} can be understood with a relatively simple random walk picture.
For this, we can expand a point in phase space in base $\beta$ such that $x = \sum_{j=1}^\infty y_j \beta^{-j}$.
The operation acts simply on this representation of $x$: $R(x) = \sum_{j=1}^\infty y_{j+1} \beta^{-j}$.
The relevant insight to connect to the random walk is that the control map multiplies $x$ by some $|a|<1$.
If we take $a = \beta^{-n}$, then we have $ax = \sum_{j=1}^\infty y_j \beta^{-(j+n)}$.
The addition of $(1-a)x_j$ in the control map serves to get us on the correct periodic orbit and can be neglected for the rest of the random walk argument (but needs careful consideration in a more rigorous treatment).
A periodic orbit will have a repeating decimal expansion $y_{j} = y_{j+N}$ for some $N\geq 1$, and when that pattern first breaks for an irrational number, that would be the ``first domain wall''.

It is then a simple matter of tracking how this first domain wall moves around.
With probability $1-p$ we apply the $\beta$-adic map which moves it 1 step to the left in this decimal expansion, and with probability $p$, we apply the control map which moves the domain wall $n$ steps to the right in the decimal expansion.
Therefore, on average, the bias in the random walk will give us a velocity of the first domain wall
\begin{equation}
    v_{\mathrm dw} = np - (1-p) = -\frac{\log(a)}{\log \beta}p - (1 - p).
\end{equation}
When this velocity switches from positive to negative, we have our controlled phase, and in fact $v_\mathrm{dw} = 0$ implies Eq.~\eqref{eq:pctrl}.
Lastly, this gives us explanation for what we see in the main text in Fig.~2.

\section{Cellular automaton formulation of the Bernoulli map}

\subsection{Chaotic map: Scrambler and orbits}
\begin{figure}
    \centering
    \includegraphics[width = 4cm]{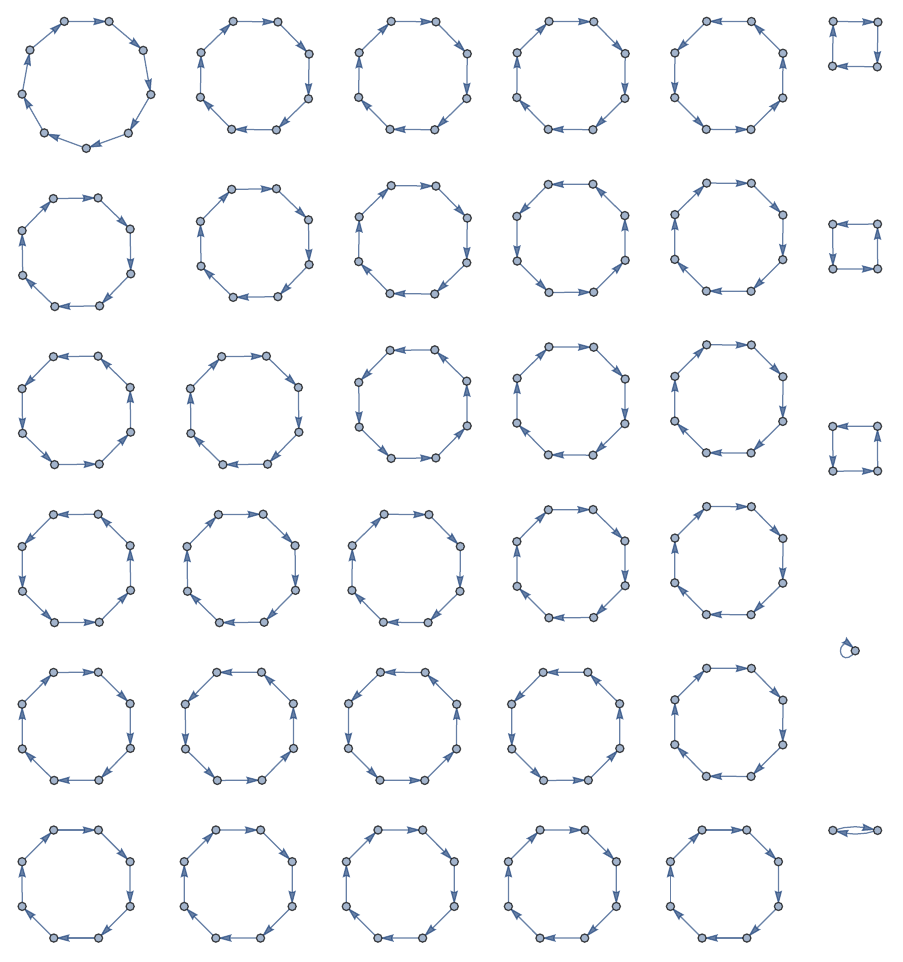} \hspace{1cm}
    \includegraphics[width = 4cm]{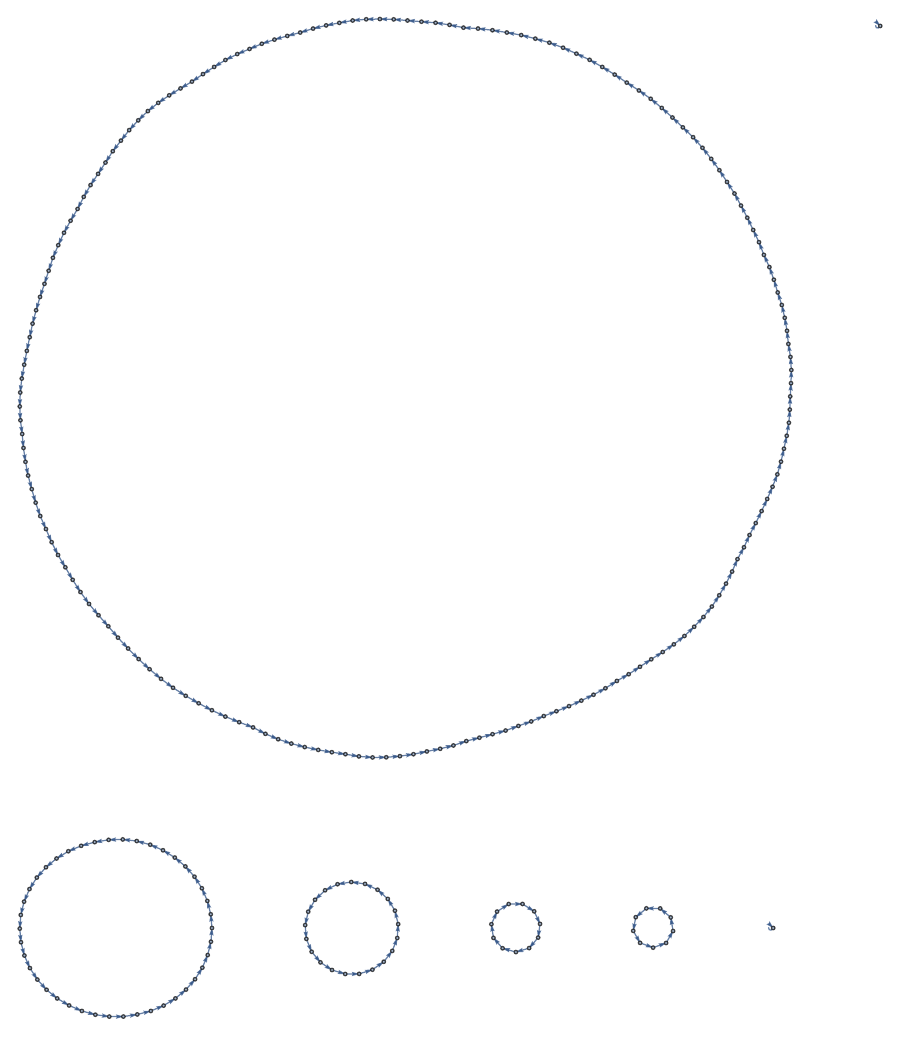} \hspace{1cm}
    \includegraphics[width = 6cm]{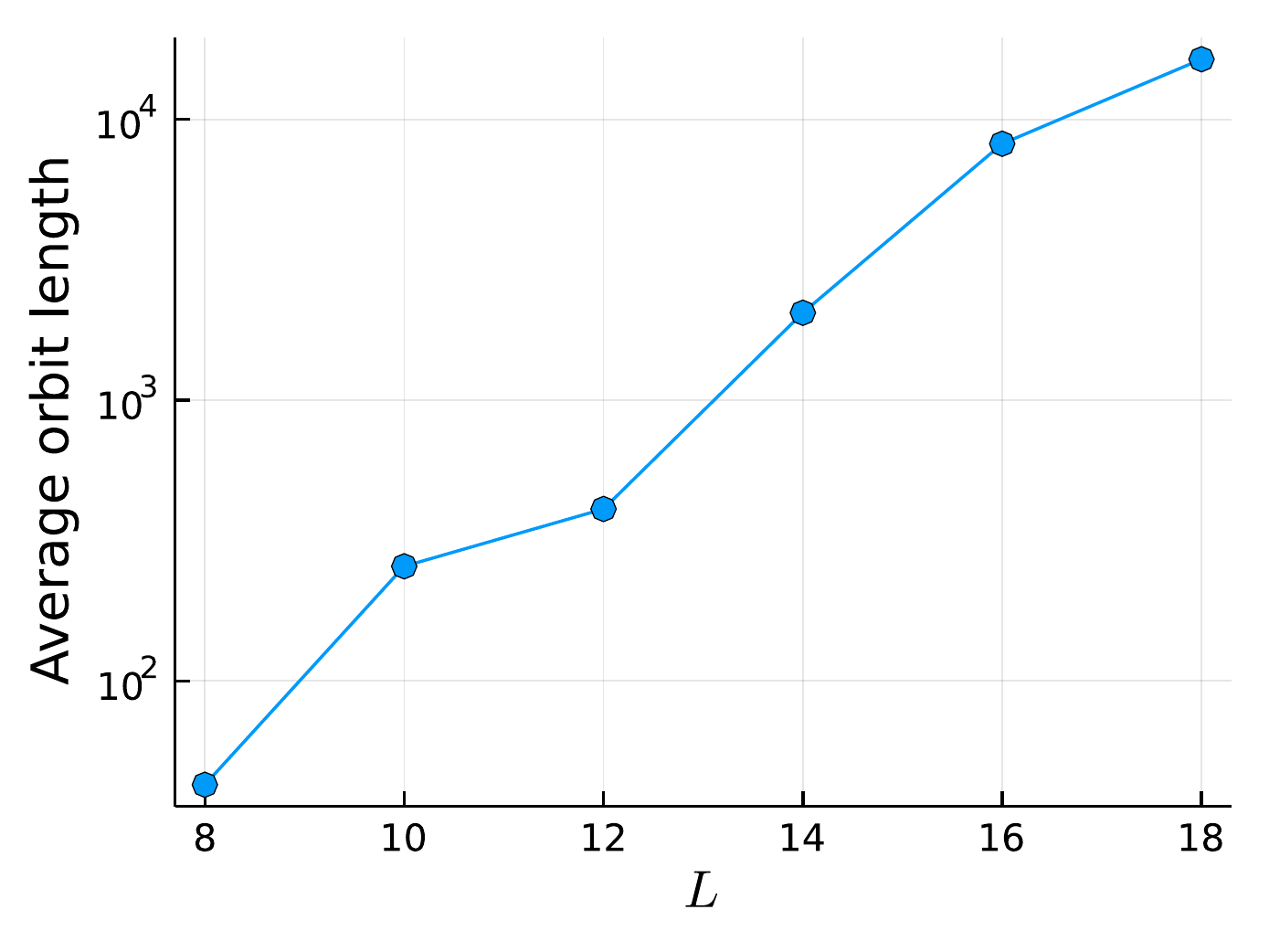}
    \caption{{\bf Cellular automaton orbits.} (left) Without the scrambler, the longest orbit is size $L$, in this case $L = 8$. (middle) with the scrambler, the orbit becomes exponential system size (this is the Bernoulli map with scrambler for $L=8$). (right) Computing the average orbit length, we find that with the scrambler it grows exponentially.}
    \label{fig:CAorbits}
\end{figure}

In this section, we provide further details on our implementation of the Bernoulli map.
As discussed in the main text, the Bernoulli map cannot simply be given by the translation operator $T$, Eq.~(3), because the resulting map has orbits of length at most $L$. The full set of orbits for a system with $L=8$ is shown in Fig.~\ref{fig:CAorbits} (left). To make the map chaotic, we append a scrambler that acts on the end of the chain corresponding to the least significant bits of the point $x\in[0,1)$.
The scrambler we use acts on the last three bits $b_{L-2} b_{L-1} b_L$ and can be represented by the cellular automaton
\begin{equation}
\label{eq:scrambler}
    S_\mathrm{cl}: \begin{cases} 111 \mapsto 001 &
    110 \mapsto 100, \\
    101 \mapsto 000, &
    100 \mapsto 011, \\
    011 \mapsto 010, &
    010 \mapsto 101, \\
    001 \mapsto 110, &
    000 \mapsto 111.
    \end{cases}
\end{equation}
The resulting chaotic map, $U^{\rm cl}_{\rm chaotic}$, is a cellular automaton whose typical orbit lengths are exponential in $L$, see Fig.~\ref{fig:CAorbits} (right). An illustration of the full set of orbits for $L=8$ is shown in Fig.~\ref{fig:CAorbits} (middle).
Note that this scrambler does not preserve the length-two orbit $0101\dots \leftrightarrow 1010\dots$ onto which we are attempting to control the system---it allows the system to leave this orbit (by an exponentially small amount in a single time step) once it reaches it. In the context of the control transition, this feature actually slightly simplifies the numerical diagnosis of the transition. However, we could avoid it by replacing $S_{\rm cl}$ with a random automaton that maps $010\leftrightarrow 101$. We have verified that the presence and location of the control transition are unaffected by this modification; our use of the scrambler~\eqref{eq:scrambler} is a choice of convenience.

\subsection{Control map: Controlled adder}

\begin{figure}
    \centering
    \includegraphics[width = 0.5\textwidth]{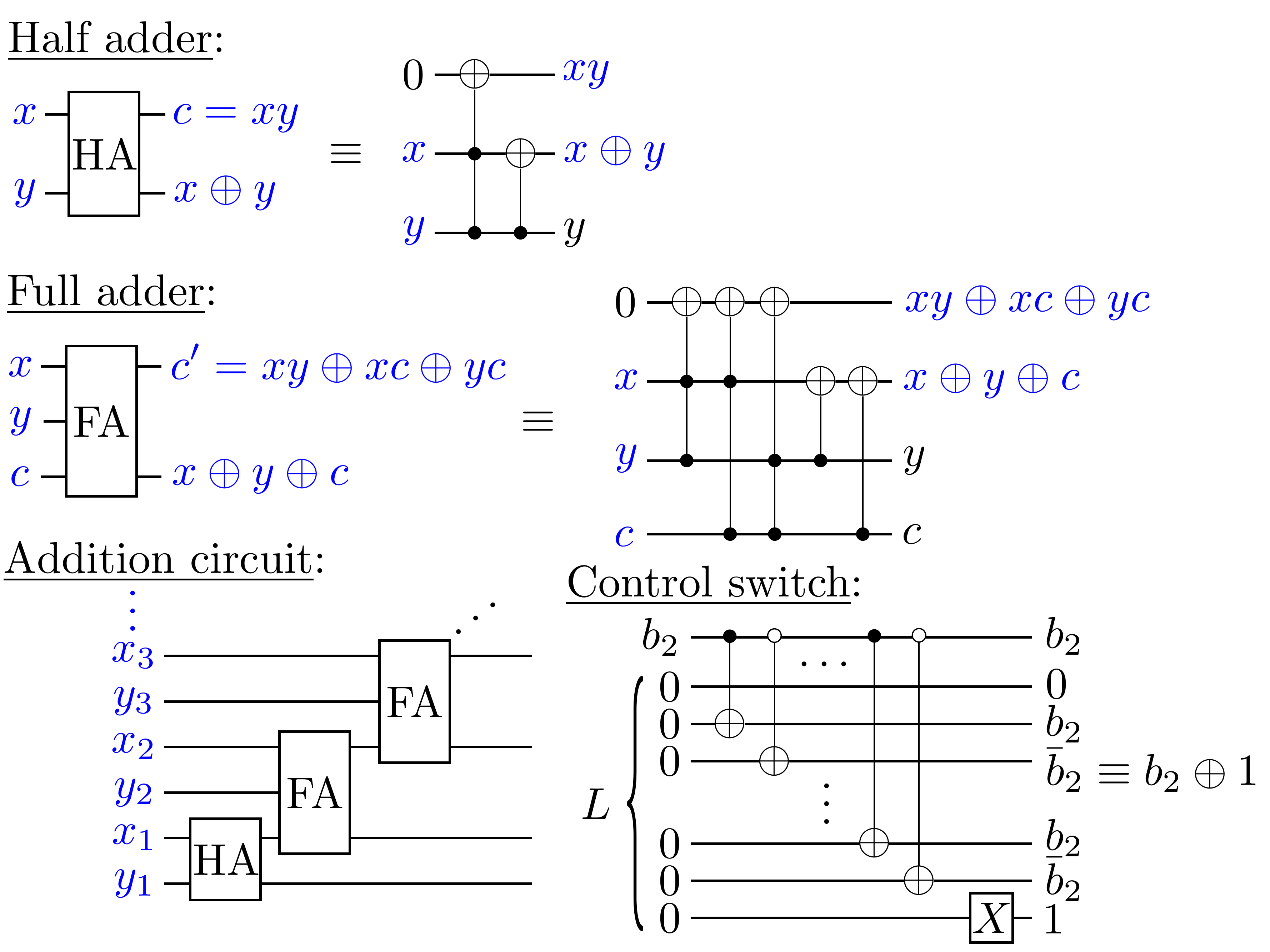}
    \caption{{\bf Ingredients for the controlled adder.} Circuits are shown for a half adder, full adder, addition circuit, and control switch, as described in the supplement text. In cases where ancilla bits are introduced, logically important bits are colored in blue; workspace and garbage bits are left black.}
    \label{fig:Adder-circuits}
\end{figure}

In this section we explain how the controlled adder $A_{\rm ctrl}$, Eq.~(7) in the main text, is constructed from local unitary gates. The adder consists of a few elementary building blocks, shown in Fig.~\ref{fig:Adder-circuits}. First, there is a half adder, which, given two bits $x$ and $y$, outputs their sum mod 2, $x\oplus y$, and a carry bit $c=xy$ which is set to $1$ if and only if $x=y=1$. The half adder is built using one Toffoli and one CNOT gate. Next, there is a full adder, which takes in two bits $x$ and $y$ as well as a carry bit $c$ (assumed to be output by a previous addition operation) and outputs the sum $x\oplus y\oplus c$ and a new carry bit $c'=xy\oplus xc\oplus yc$ which is $1$ if and only if two or more of $x,y,$ and $c$ are $1$. This can be achieved with three Toffolis and two CNOTs. In order for these operations to be logically reversible (i.e., compatible with a unitary implementation), we need to augment the bits storing logical information (colored in blue in the figure) with input ancilla bits, which provide workspace for the circuit, and output garbage bits, whose values can be erased after performing the operation. The addition circuit is then obtained by forming a staircase of half and full adders, as shown at the bottom left of Fig.~\ref{fig:Adder-circuits} suppressing the ancilla and garbage bits for clarity.

\begin{figure}
    \centering
    \includegraphics[width = 0.5\textwidth]{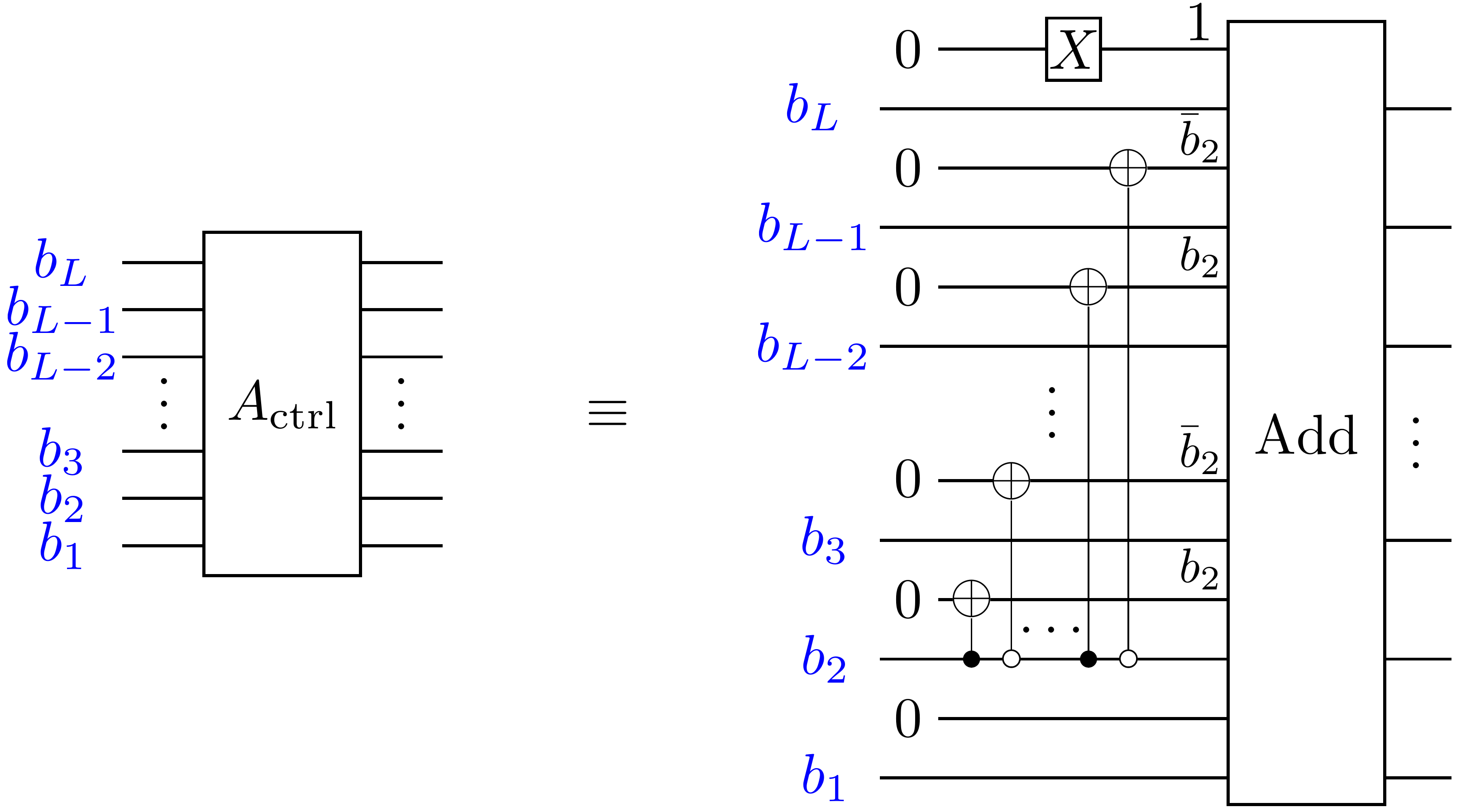}
    \caption{{\bf Controlled adder circuit.} The controlled adder $A_{\rm ctrl}$ is constructed by applying the control switch circuit from the bottom right of Fig.~\ref{fig:Adder-circuits} and feeding the result into the addition circuit from the bottom left of Fig.~\ref{fig:Adder-circuits}, along with the bitstring $b_1b_2\dots b_L$ that describes the state of the system (colored blue).}
    \label{fig:Actrl}
\end{figure}

To construct the controlled addition circuit we need to prepend to the adder described above with a ``control switch" circuit, whose purpose is to determine which binary fraction to add to the input bitstring $0.b_1b_2\dots b_L$. As noted in the main text, the fraction to be added is conditioned on the value of $b_2$; from Eq.~(7), we see that it is given by $0.0 b_2 \bar{b}_2 \dots b_2 \bar{b}_2 1$, where $\bar{b}_2=b_2\oplus 1$ is the logical complement of $b_2$. This bitstring can be constructed from a register of $L$ ancillas initialized in $0$ using CNOT gates as shown at the bottom right of Fig.~\ref{fig:Adder-circuits}. The controlled adder $A_{\rm ctrl}$ then takes the form shown in Fig.~\ref{fig:Actrl}.

\section{Krauss operator description}

For the description in terms of quantum channels, we use the notation developed in the main text [specifically Eqs.~(3-7)].

There are three Krauss operators:

\begin{equation}
    \begin{aligned}
     K_U & \equiv \sqrt{1-p} U_{\mathrm{ chaotic}}^{\mathrm{qm}}, \\
     K_C^0 & \equiv \sqrt{p} A_{\mathrm{ctrl}}T^{-1} P_L^0, \\
     K_C^1 & \equiv \sqrt{p} A_{\mathrm{ctrl}}T^{-1} X_L P_L^1.
    \end{aligned}
\end{equation}
$K_U$ describes the Bernoulli map applied with probability $1-p$ and $K_C^j$ represents the control map along with its two outcomes for the measurement: 0 and 1.
Unitary control is applied, just as in the main text, when the outcome of measuring bit $L$ is 1.
These satisfy $ K_U^\dagger K_U + (K_C^0)^\dagger K_C^0 + (K_C^1)^\dagger K_C^1= 1$, and the \emph{average} density matrix satisfies is progressed from time $t$ to $t+1$ via
\begin{equation}
    \rho_{t+1} = K_U \rho_t K_U^\dagger + K_C^0 \rho_t (K_C^0)^\dagger + K_C^1 \rho_t (K_C^1)^\dagger.
\end{equation}
Note that this is already distinct from the \emph{individual trajectories} as we numerically simulate (which are characterized by products of chaotic evolution and the control operation).
Entanglement data (ancilla, half-cut, etc.) cannot be directly computed with this expression, that still requires nonlinear observables in the density matrix of individual trajectories.

For purposes of applying the scrambler, we note that the density matrix can be written in tensor form
\begin{equation}
\begin{array}{c}
    \includegraphics{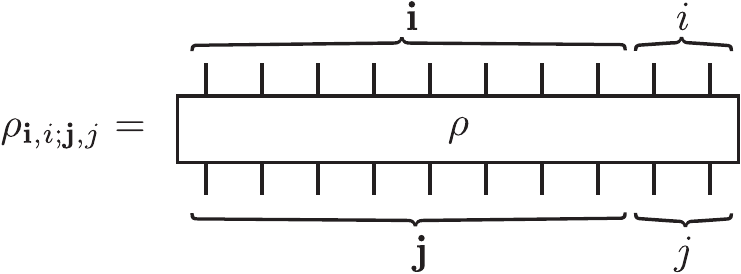}
\end{array}
\end{equation}
where each line represents a qubit (1 to $L$, reading left to right).
The average of the Krauss map results reveal a connection between the \emph{quantum} model and \emph{classical probability distributions} due to
\begin{equation}
    \int dS \,\sum_{k,l=1}^4 S_{ik} \rho_{\mathbf{i}, k; \mathbf{j},l} (S^\dagger)_{lj} = \frac{1}{4} \delta_{ij} \sum_{k=1}^4 \rho_{\mathbf i, k; \mathbf{j}, k},
\end{equation}
where $dS$ is the Haar measure over dimension 4 unitaries.
Furthermore, if we put the density matrix through a dephasing channel to suppress off-diagonal contributions, we can write
\begin{equation}
    \braket{ \braket{\rho_t}} = \sum_{\mathbf b} p_t(\mathbf b) \ket{\mathbf b}\! \bra{\mathbf b},
\end{equation}
where all $p_t(\mathbf b) \geq 0$ and $\mathbf b = (b_1, b_2, \ldots, b_L)$ are qubit $Z$-states.
If we track the time evolution of $\braket{\braket{\rho_t}}$ (indicating an average over Haar distributed $S$ and a dephasing channel on $\rho_t$), then we get

\begin{equation}
\begin{aligned}
    \braket{\braket{K_U \rho_t K_U^\dagger}} & = \frac14 (1-p) \sum_{\mathbf b} \left(\sum_{b_1,b_L} p_t(\mathbf b) \right) \ket{b_2 b_3 \cdots b_{L-1}}\bra{b_2 b_3 \cdots b_{L-1}} \otimes \mathbf I, \\
    \braket{\braket{K_C^b \rho_t (K_c^b)^\dagger}} & = p \sum_{\mathbf b | b_L = b, b_1 = 0} p_t(\mathbf b) \ket{00 b_2 \ldots b_{L-1} + 001010\cdots 011}\bra{00 b_2 \ldots b_{L-1} + 001010\cdots 011}\\
    & \quad  + p \sum_{\mathbf b | b_L = b, b_1 = 1} p_t(\mathbf b) \ket{01 b_2 \ldots b_{L-1} + 01010\cdots 101}\bra{01 b_2 \ldots b_{L-1} + 01010\cdots 101}.
\end{aligned}
\end{equation}

In order to proceed, we want to find an expression for $p_{t+1}(\mathbf b) = V\rho_t(\mathbf b)$ for an operator $V$.
We make a few definitions, using Eqs.~\eqref{eq:supp_bern} and \eqref{eq:supp_ctrl} for notational help.
First
\begin{equation}
C(\mathbf b)  = 0b_1b_2\cdots b_{L-1} + \begin{cases}
  001010\cdots 011, & \text{if $b_1 = 0$}, \\
  010101 \cdots 101, & \text{if $b_1 = 1$}.
\end{cases}
\end{equation}
Notably, the inverse map $C^{-1}(\mathbf b)$, when it exists, is a set of two possibilities $C^{-1}(\mathbf b) = \{ \tilde b_1 \cdots \tilde b_{L-1} 0,\tilde b_1 \cdots \tilde b_{L-1} 1 \}$, for $\tilde b_j$ which satisfy $C(\tilde b_1 \cdots \tilde b_{L-1} B) = \mathbf b$.
This lets us write
$$
    \sum_b \braket{\braket{K_C^b \rho_t (K_c^b)^\dagger}}  = p \sum_{\mathbf b} \left(\sum_{\tilde{\mathbf b} \in C^{-1}(\mathbf b)} p_t( \tilde {\mathbf b}) \right) \ket{\mathbf b}\bra{\mathbf b},
$$
and furthermore
$$
    \braket{\braket{K_U \rho_t K_U^\dagger}}  = \frac14 (1-p) \sum_{\mathbf b} \left(\sum_{\tilde b_{L-1},\tilde b_L} p_t(b_{L} b_1 b_2 \cdots b_{L-2} b_{L-1}) \right) \ket{\mathbf b}\bra{\mathbf b} .
$$
To connect this last expression with the classical map, call the operation $R(\mathbf b) = R(b_1 \cdots b_L) = b_2 b_3 \cdots b_{L-2} 0 0$ which we combine with a scrambling operation $S$ which with equal probability takes $00$ to $\{00,01,10,11\}$.
In this way, $S^{-1}(b_1 \cdots b_{L-2} b_{L-1} b_L) = b_2 \cdots b_{L-2} 00$ and $(S \circ R)^{-1}(b_1 \cdots b_L) = \{\tilde b_{L} b_1 \cdots b_{L-2} \tilde b_{L-1}\quad \text{for}\quad  \tilde b_L, \tilde b_{L-1} \in \{0,1\}\} $, and we therefore have
$$
    \braket{\braket{K_U \rho_t K_U^\dagger}}  = \frac14(1-p) \sum_{\mathbf b} \left(\sum_{\tilde b \in (S\circ R)^{-1}(\mathbf b)} p_t(\tilde{\mathbf b}) \right) \ket{\mathbf b}\bra{\mathbf b} .
$$

Taken together, we have an equation for the time evolution of probability distributions
\begin{equation}
    p_{t+1}(\mathbf b) = V\rho_t(\mathbf b) = \frac1{4}(1-p)\sum_{\tilde b \in (S\circ R)^{-1}(\mathbf b)} p_t(\tilde{\mathbf b})  + p \sum_{\tilde{\mathbf b} \in C^{-1}(\mathbf b)} p_t( \tilde {\mathbf b}).
\end{equation}
And finally, we need to be careful about taking the thermodynamic limit and relating it back to the Bernoulli map.
In both cases the last bit $b_L$ is irrelevant and two bit strings $b_1 \cdots b_{L-1} 0$ and $b_1 \cdots b_{L-1} 1$ both will map to the same $x$ point in $[0,1)$, yet they are summed over above.
Therefore, we will identify these points such that $p_t(b_1 b_2 \cdots b_{L-1} 0 ) + p_t(b_1 b_2 \cdots b_{L-1} 1) \rightarrow p_t(x)$ where $x = 0.b_1 b_2\cdots$.
In this limit, we can write the above expression
$$
p_{t+1}(x) = Vp_t(x) = \frac1{2} (1-p) \sum_{\tilde x \in R^{-1}(x)} p_t(\tilde x) + p \sum_{\tilde x \in C^{-1}( x) } p_t(\tilde x),
$$
where we have lost the scrambler $S$ in the thermodynamic limit.
Note that this defines the Frobenius-Perron operator (Eq.~\ref{eq:FPtotal}) defined in, for instance, Ref.~\cite{antoniou_probabilistic_1996} for $\beta = 2$ and $a = 2^{-1}$, proving that Krauss evolution of our density matrix is equivalent to classical evolution of probability distribution functions when we average over the Haar unitaries and use a dephasing channel.

\section{Additional data for the quantum model}

\subsection{Order parameter and entanglement saturation}

\begin{figure}
    \centering
    \includegraphics[width=0.8\textwidth]{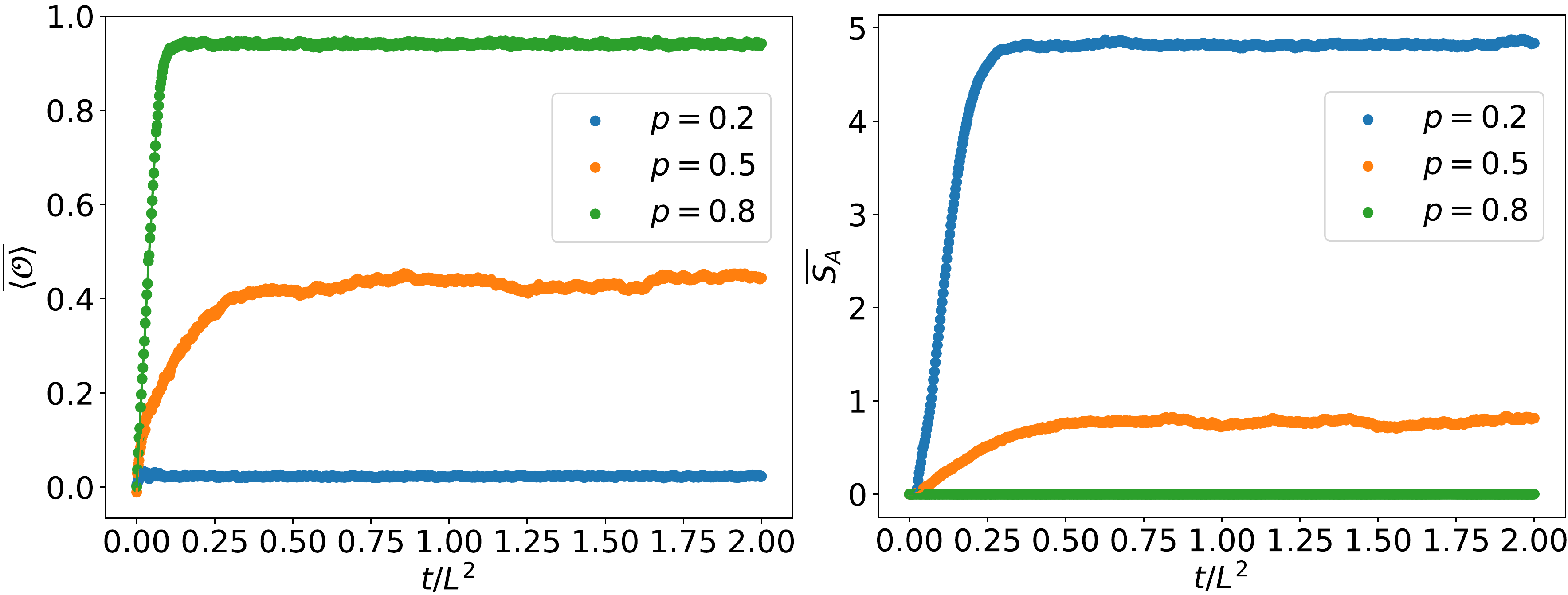}
    \caption{{\bf Order parameter and entanglement dynamics.} Dynamics of the realization-averaged order parameter $\overline{\braket{\mathcal O}}$ for $L=20$ (left) and half-cut entanglement entropy $\overline{S_A}$ for $L=18$ (right) at representative values of $p=0.2, 0.5,$ and $0.8$. For all values of $p$ considered, the order parameter and entanglement reach a steady state well before the end of the simulation time window of $2L^2$. (Error bars and number of realizations are as in main text.)}
    \label{fig:Saturation}
\end{figure}

In Fig.~\ref{fig:Saturation}, we plot the dynamics of $\overline{\braket{\mathcal O}}$ for $L=20$ and $\overline{S_A}$ for $L=18$ (averaged, as in the main text, over $1000$ realizations) out to a time $t=2L^2$. For all values of $p$ we considered ($p=0.2, 0.5,$ and $0.8$ are shown as representative values below, at, and above the control transition, respectively), the order parameter and half-cut entanglement entropy have saturated well before the final time in the simulation window. This justifies our use of times of order $L^2$ to probe properties of the steady state.

\subsection{System-size scaling of entanglement dynamics in the uncontrolled phase}

\begin{figure}
    \centering
    \includegraphics[width=0.8\textwidth]{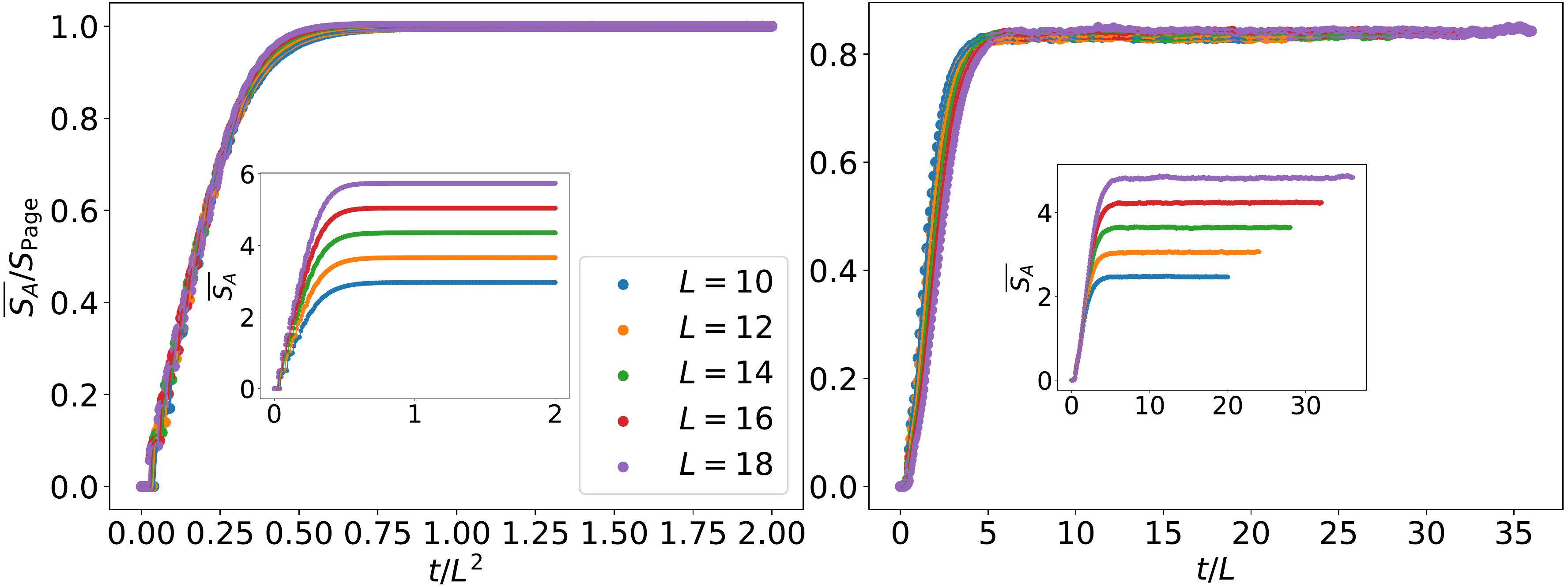}
    \caption{{\bf System-size dependence of entanglement dynamics at small $p$.} (left) Realization averaged entanglement entropy $\overline{S_A}$ in units of $S_{\rm Page}$ as a function of $t/L^2$ for $L=10,\dots,18$ at $p=0$. Inset: Same as main panel except without rescaling the $y$-axis by the Page value. (right) Realization averaged entanglement entropy $\overline{S_A}$ in units of $S_{\rm Page}$ as a function of $t/L$ for $L=10,\dots,18$ at $p=0.2$. Inset: Same as main panel except without rescaling the $y$-axis by the Page value. (Error bars and number of realizations are as in main text.) }
    \label{fig:TimeScaling}
\end{figure}

In this section, we corroborate two statements made in the main text regarding the system-size dependence of the entanglement dynamics in the uncontrolled phase. The first statement, made below Eq.~(4), is that the entanglement generated by the chaotic unitary $U^{\rm qm}_{\rm chaotic}$ saturates to a volume-law value in an $O(L^2)$ time. Fig.~\ref{fig:TimeScaling}(left) plots the realization-averaged entanglement entropy $\overline{S_A}$ against $t/L^2$ at $p=0$ for $L=10,\dots,18$. When $\overline{S_A}$ is rescaled by the Page value, $S_{\rm Page}=\frac{L}{2}\ln 2 - \frac{1}{2}$, for the entanglement entropy of a random state~\cite{Page93}, we find that the dynamics collapse onto a single curve. This indicates that $U^{\rm qm}_{\rm chaotic}$ generates volume-law entanglement in an $O(L^2)$ time, as claimed in the main text.

The second statement, made immediately before the \textit{Discussion and outlook} section, is that volume-law entanglement can only develop at the transition if the FDW ``sticks" to the left edge of the chain for at least an $O(L)$ time. To simulate this scenario, we consider the realization-averaged dynamics at $p=0.2$. Since a randomly chosen CB state has, with high probability, a FDW located near the left end of the chain, the average over realizations at $p=0.2$ probes what happens when the FDW is initialized near the left edge of the chain. We are then interested in the timescale for entanglement saturation starting from these initial conditions. Fig.~\ref{fig:TimeScaling}(right) plots $\overline{S_A}/S_{\rm Page}$ against $t/L$ at $p=0.2$ for $L=10,\dots,18$. The rescaled data show a slight rightward drift of the entanglement saturation time as system size increases. This means that the entanglement saturation timescale can be lower-bounded by an $O(L)$ value. Here, the low value of $p=0.2$ is chosen to encourage the FDW to stick to the left edge of the chain. However, at the transition $p_{\rm ctrl}\approx 0.5$, the dynamics of the FDW becomes an unbiased random walk. Thus, for the FDW to stick to the left edge of the chain for a time of at least $O(L)$ requires the random walker to step to the left $O(L)$ times, and the probability of this happening is exponentially small in $L$.

\subsection{Fluctuations at the transition}

\begin{figure}
    \centering
    \includegraphics[width=0.8\textwidth]{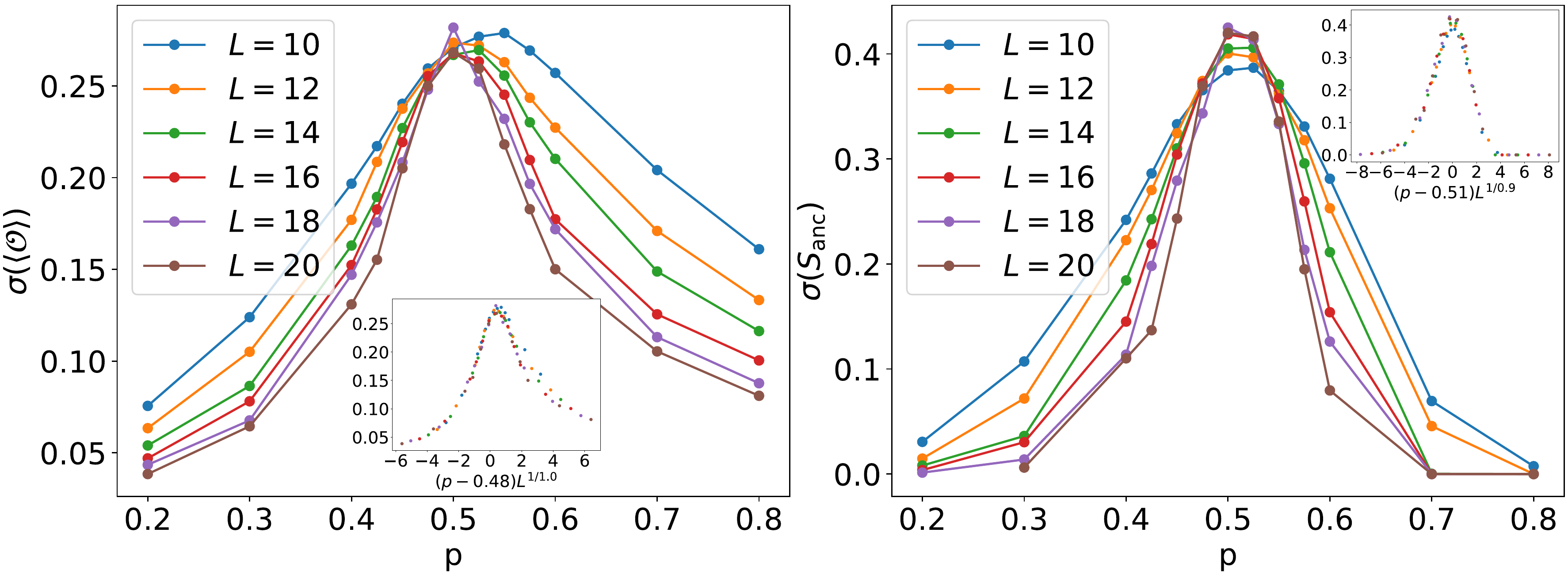}
    \caption{{\bf Fluctuations.} Standard deviation over realizations of the order parameter $\braket{\mathcal O}$ and the ancilla entanglement entropy $S_{\rm anc}$, using the same datasets that produced Fig.~3 in the main text. Insets show data collapse assuming the respective values of $\nu$ and $p_{\rm ctrl}$ determined in the main text.}
    \label{fig:Fluctuations}
\end{figure}

The location of the control transition can also be probed by tracking the fluctuations of observables as a function of $p$ and $L$. In Fig.~\ref{fig:Fluctuations}, we show the standard deviation over realizations of the observables considered in Fig.~3 of the main text. For both the order parameter $\braket{\mathcal O}$ and the ancilla entanglement entropy $S_{\rm anc}$, the fluctuations exhibit pronounced peaks near $p=0.5$. For $\braket{\mathcal O}$, the peak appears to drift towards $p=0.5$ and become sharper with increasing system size $L$. For $S_{\rm anc}$, the peak exhibits less drift in its position but still sharpens with increasing $L$. Both datasets also exhibit a near collapse assuming the respective values of $\nu$ and $p_{\rm ctrl}$ determined in the main text.

\end{document}